# The Principle of Synergy and Isomorphic Units

# Edgar Paternina

Electrical engineer Author of **Physics and The Principle of Synergy**, published in CD ROM in English and Spanish at Amazon.com, on which this paper is based

Contact:epaterni@epm.net.co

### Abstract

A solution to the part and whole problem is presented in this paper by using a complex mathematical representation that permits to define the Holon concept as a unit that remains itself in spite of complex operations such as integration and derivation. This can be done because of the remarkable isomorphic property of Euler Relation. We can then define a domain independent of the observer and the object, as within it, the object is embedded. We will then be able to have a Quantum Mechanics solution without the "observer drawback", as Karl R. Popper tried to find all his life but from the philosophical point of view and which was Einstein main concern about QM too. A unit that has always similar or identical structure or form, despite even complex operations such as integration and derivation, is the ideal unit for the new sciences of complexity or just the systems sciences too, where structure or form, wholeness, organization, and complexity are main requirements. A table for validating the results obtained is presented in case of the pendulum formula.

## Complex Numbers and Isomorphic Units

In his **General System Theory**, Ludwig von Bertalanffy wrote:

*Reality, in the modern conception, appears as a tremendous hierarchical order of organized entities, leading, in a superposition of many levels, from the physical and chemical to biological and sociological systems. Unity of Science is granted, not by a utopian reduction of all sciences to physics and chemistry, but by the structural uniformities of the different levels of reality.*

Those structural uniformities or isomorphisms in different levels of reality are the main concern of this paper, and its main aim will be to present a new way of "seeing" reality by means of some *isomorphic units, or co-variant units*, so to speak, units in which the form is one important attribute as well. A basic recurrent design or pattern that can be used to interpret and explain those problems where dynamic interactions or an organized complexity appear.

But the problem of form appeared in classical physics too but precisely in those fields where the **field concept** was unavoidable. The magnetic field problem, whose existence can even be felt by putting two permanent magnets near by, is really one of those problems of nature that after all were hiding a great mystery. With the magnet we can not only present theoretical examples of those three basic fundamental attributes that are the basic to an isomorphic unit, but the magnetic field to be well represented, from the mathematical point of view, we must also use a complex number mathematical symbolism.

Those three basic attributes are: wholeness, oneness and openness.

The wholeness attribute can be seen easily in the case of a magnet, where each magnet split from another magnet is precisely a whole new magnet. **This new "part" that came from an old whole is a whole too**.

To obtain a whole from another whole is like to obtain a son from a father, or an object from the instanciation of a class and in this same sense Ken Wilber wrote *To be a part of a larger whole means that the whole supplies a principle(or some sort of glue) not found in the isolated parts alone, and this principle allows the parts to join, to link together, to have something in common, to be connected, in ways that they simply could not be on their own...When it is said that "the whole is greater than the sum of its parts," the "greater" means "hierarchy"...This is why "hierarchy" and "wholeness·" are often uttered in the same sentence*

But associated with this wholeness attribute is that binary or dual aspect of reality, where we have always two "opposites" or more appropriately, complementary or polar entities within a comprehensive whole, or just a unit that transcends duality, the oneness attribute, that can be found physically in a magnet. The within and the without some sort of dynamic structure embedded in a unit that cannot be split in its two components. Structures that perform well in a changing environment must be capable to reflect the without, they must have, as it were, a ***storing capacity*** to reflect that without.

But the magnet has also another remarkable attribute associated with form, with the environment or interface or some sort of a medium to separate an internal milieu from an external environment, as it were, a field concept, or the openness attribute. At this point is important to recall that these three attributes cannot be considered on their own. They are linked together by some sort of glue. This is the main characteristic of this new way of "seeing" reality in which we must always have in mind that glueing principle, we will name the Principle of Synergy. But at this point of reasoning we are taken to think in open systems, as those *systems capable of exchanging with the environment.*

Prejudices normally come from ways of "seeing" reality that had generated by themselves some sort of secure stance in one of those realms. Facts, data or immediate experiences can be obtained in all those three realms, of science, philosophy and ontology and in every one of those realms we find in fact an actual practice, some sort of methodology to obtain facts. But the data obtained in every case is limited by the language used and with objects of study that are obtained. Those that practice philosophy or ontology normally have a great prejudice against a mathematical language, which is some sort of language that permits the practicians to avoid any ambiguity, by obtaining with it precise definitions. But on the other side those that practice philosophy and ontology were always right in the sense that a language whose main aim were to avoid any ambiguity or uncertainty was not a good language to interpret reality after all. That ambiguity or uncertainty can be found there where we find those three attributes, that has to do with open systems and in general with dynamism or generation of forms: wholeness, oneness and openness.

For representing mathematically such kind of problems, it is necessary to use a language that permits to define a unit embedded in a comprehensive whole, environment or dynamic structure, and which can include also a radical duality or polarity, which can be generated, precisely because that inherent coincidentia oppositorum or tension, as it were, a field, that makes by definition that unit an open system. Complex numbers have the capacity not only to represent that binary or duality aspect of reality or the chance to have two polarities included in one unit, but also the nondual or wholeness attribute, we have related with that capacity to generate a new whole. The openness attribute is related with a capacity to generate a field that is concomitant with those entities we can named holonic -to use a term coined by Ken Wilber- such as magnetic entities, electrons, linguistic signs, cells, life, mind and beings in general.

But complex numbers were born when trying to solve the simple algebraic equation

$$x^2 + 1 = 0$$

$$x^2 = -1$$

$$x = \mathrm{Sqr}(-1) = \mathbf{J}$$

where as a solution we have the square root of a negative one or the radical unit **J**.

From the point of view of "real" number perspective it does not exist a solution for this simple problem and as so it was necessary to make a paradigm extension or paradigm shift, and to define a new type of more general numbers to solve the problem. The solution **J** was named "imaginary" by Descartes for the first time and since then, complex numbers remained as some sort of a strange mathematical tool. It was Leonard Euler in 1745, the one, that finally found a mathematical symbol for representing that new entity that included both kinds of numbers:

- those named real and which we will term nondual for reasons we will see later, and

- those named "imaginary", we will name dual on the other side for apparent reasons too.

Problems of growth has been associated from the point of view of a mathematical representation with the number epsilon or euler number since a long time ago. And it was by studying infinite series, that Euler found that entity, that not only could represent those two kinds of numbers, nondual and duals, but also it included those cyclical waveform, sine and cosine, that occur so frequently in nature wherever we have cyclical phenomena. But the most important, the most cogen argument to use this kind of new numbers, that have been used by Electrical Engineering since Oliver Heaviside and Steinmetz introduced them at the end of the 19th century to solve alternating electrical current circuits, is precisely, its inherent isomorphic property, that permits them to make, as it were, co-variant representations, or to make simpler complex operations.

Evidently such a useful mathematical entity, was adequate to represent dynamic realities, and it was used for the first time for representing electromagnetic fields at the end of the 19th century, and after that, vectors were born in physics, but then they are taught, in general, without making any references to this complex number origin. Up to that moment complex numbers had not been used in practical cases, and as so it justifies why the complex plane was delayed 100 years from its real birth at the end of the 18th century.

A unit that has always similar or identical structure or form, despite even complex operations such as integration and derivation, is the ideal unit for the new sciences of complexity or just the systems sciences too, where structure or form, wholeness, organization, and complexity are main requirements. But another important point is that it can also be used to define then a Basic Unit System concept in which uncertainty is included, and as so open systems.

Classical physics had as it main aim to resolve natural phenomena into a play of elementary units, as it were, to resolve those phenomena into their parts, isolated parts, I mean, so the concept of particle was always the starting point of the whole framework. But that part or particle needed to be considered as an isolated entity, that is, as a closed system, in which there were no interactions at all with the environment. This ideal model to represent reality was so restrictive that it definitely failed, the way we all know.

An adequate framework for representing reality, the whole reality, must be complex, in the sense, that it must not only include, the dual-logical-nature of reality, but also it must include, that another aspect related with form, wholeness and oneness. The form, the structure must be co-variant, that is, it must remain the same in spite of a progressive modification of that same structure. In this way adaptation or changeness as a fundamental process can give us persistent properties for that structure or just despite complex operations such as integration and derivation done upon that structure. But there is another more important aspect to recall, and it is the need not to reduce uncertainty, in certain cases, but to be able just to manage it. And here we come across with the fundamental problem of open and closed systems. When we have an open system, in general, that uncertainty is an essential part of the problem, as it were, of its openness attribute, and in this sense that uncertainty cannot be reduced unless we close the system, so we determine its state completely defining then ideal objects of study, that can fail in real cases. The interactions of a closed system are reduced almost to zero and then the system becomes a static system and not precisely as a steady state system. The so-called-new-sciences-of-complexity, with the seminal work of John H. Holland, **Adaptation in Natural and artificial systems** has as a main drawback precisely this tendency to eliminate uncertainty, reducing it to unprecisions or just by trying to close the system under consideration, so it is not strange to find in that school of science a tendency not abandon the Second Law of Thermodynamics as the main principle. But open systems and that second law are in some sense "incompatible" if we establish a hierarchical framework in which, that second law is just a special case of the behavior of an open system.

## Euler Relation and Its Isomorphic Properties

Up to this moment we have been seeing the emergence of a new concept of unit that includes in its mathematical representation the dual and nondual nature of reality, as it were, a relationship between the part and the whole or just a unit that is a whole and a part at the same time or a Holon as Ken Wilber named it. And what we aim with this paper is to show that Euler Relation, is precisely the mathematical symbol necessary to make an adequate representation of that whole/part entity we have named a Bus:

$$e^{J(\emptyset)} = \cos(\emptyset) + J \sin(\emptyset)$$

By asigning values to Ø, from Ø = 0 to Ø = 90 degrees, we obtain an horizontal and a vertical line respectively, as it were, the complex plane, which can be "seen" as a fifth sphere of reality or a totality that can be used to represent or contain the four dimensional space-time continuum. It can also be seen as a mathematical representation of the domain of Form, in the same line of that domain imagined by the perennial philosophy with Plato. Karl R. Popper in the intent to transcend dualism envisioned this domain as a "third world", independent from mind or the subject and the object. Karl R. Popper main concern in Objective Knowledge was precisely to avoid what he called an essentialist explanation by introducing this "third world", as a world independent both from the object and the subject. But only through a mathematical representation we can avoid any semantic pitfall. In figure we can see two systems **S** and **S´** in interrelation, and apart each other an angle, being the whole domain of representation the complex plane.

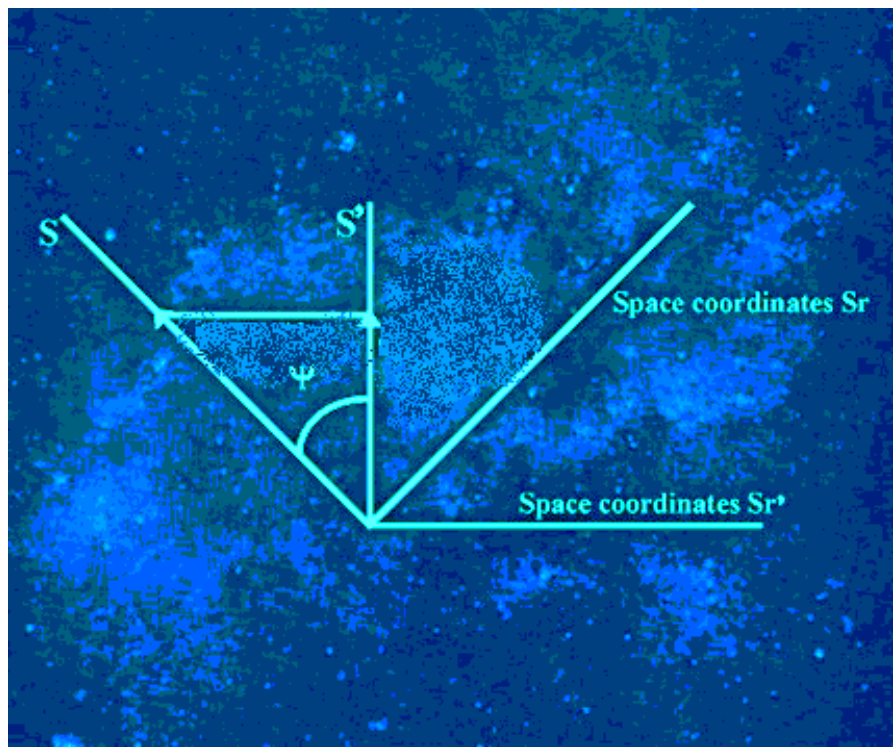

And in general we can then envision Ø acquiring all values, from zero to 360 degrees, or just, some sort of a clock pointer, or a vector rotating about an origin at a given frequency, where Ø = wt and w = 2¶f

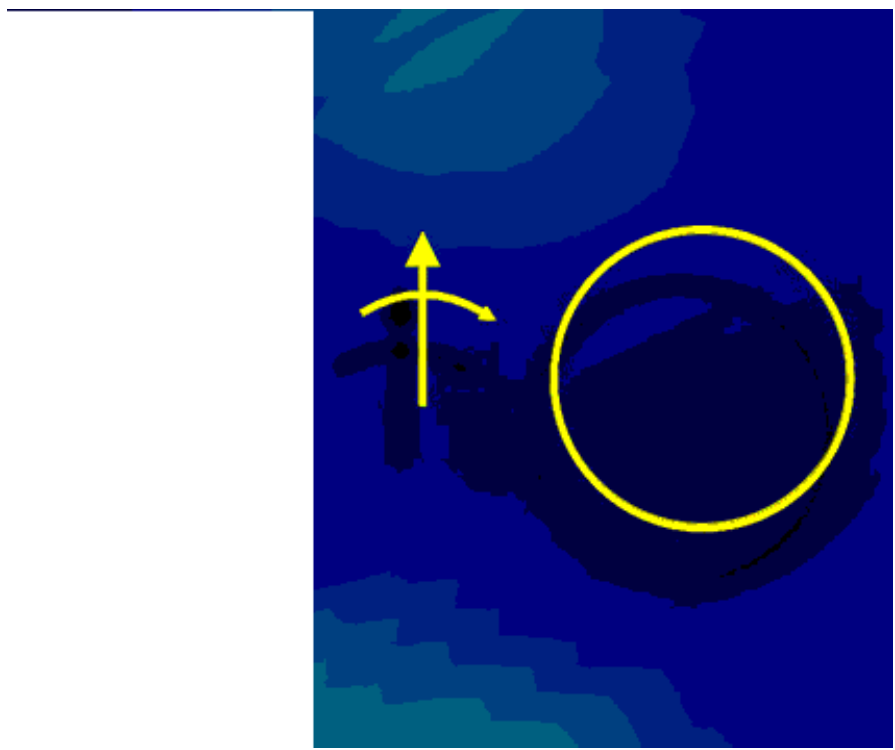

A vector rotating around an origin or a point, is what we term the **centerness** attribute of Euler Relation, that has to do with its natural cyclical behavior and we could say nature loves the cyclical waveform behavior, as we find it everywhere, from the motions of the stars and the planets, to the tides on earth, our heartbeats and our psycological states and even that motion that has always exerted such a fascination upon human mind, and I mean the pendulum motion. In Euler relation we have two types of cyclical waves separated by the radical J, and what this means is that those two type of cyclical waves are very different in nature, and this is what we are going to show in the following.

In Euler Relation we then have a:

Dual, symmetric or binary component that can be represented with the sine function:

$$(-) = \sin(-\phi) \qquad (+) = \sin(\phi)$$

and aNondual or Oneness component that can be represented with the cosine function:

$$\cos(\emptyset) = \cos(-\emptyset)$$

We must recall that the dual component or that component associated with the sine function changes with changing the sign of the angle Ø, and that the nondual component or cosine function remains the same with changing that sign angle, so we have in this unit those two requisites we pointed out at the beginning of this paper necessary for representing reality. But this mathematical process of changing sign is associated with changing the rotation sense, counterclockwise or clockwise as we will see later, so it has to do with a very general sense of rotation of the whole structure.

Historically Euler Relation was associated with the problem of the infinite series:

$$e^x = 1 + x + x^2/2! + x^3/3! + ...$$

where by replacing

$$X = J*\emptyset$$

we obtain finally Euler Relation by separating those terms that are affected by J from the others, obtaining the nondual or cosine expresion and the dual affected by J, or sine expression.

This infinite series can be called Euler Relation, where, e = 2.71828. But the main point to notice at this very moment, is the cosine and sine nature of the two components separated by **J**.

*Cosine or nondual nature of Euler Relation*

From Euler relation we obtain the cosine function represented by

$$\mathbf{Cos\ (wt)\ =\ (e^{J(wt)})/2 + (e^{-J(wt)})/2}$$

so the cosine function is expressed as the sum of two vectors rotating in opposite directions, one of them in counterclockwise or positive direction at an angular velocity w, and the second one in the clockwise or negative direction at an equal angular velocity w and as the vector rotate the two dual components cancel each other, and as so the sum is a purely real or nondual vector, nondual because we cannot obtain the opposite by changing the sign of the angle. The axis of the cosine function is in fact that axis not affected by J, the so called real axis, we have named the nondual axis. At this point it is important to recall the radical duality, expressed in that inherent polarity or tension of those two vectors rotating in opposite directions but in a comprehensive whole, or just the complex plane.

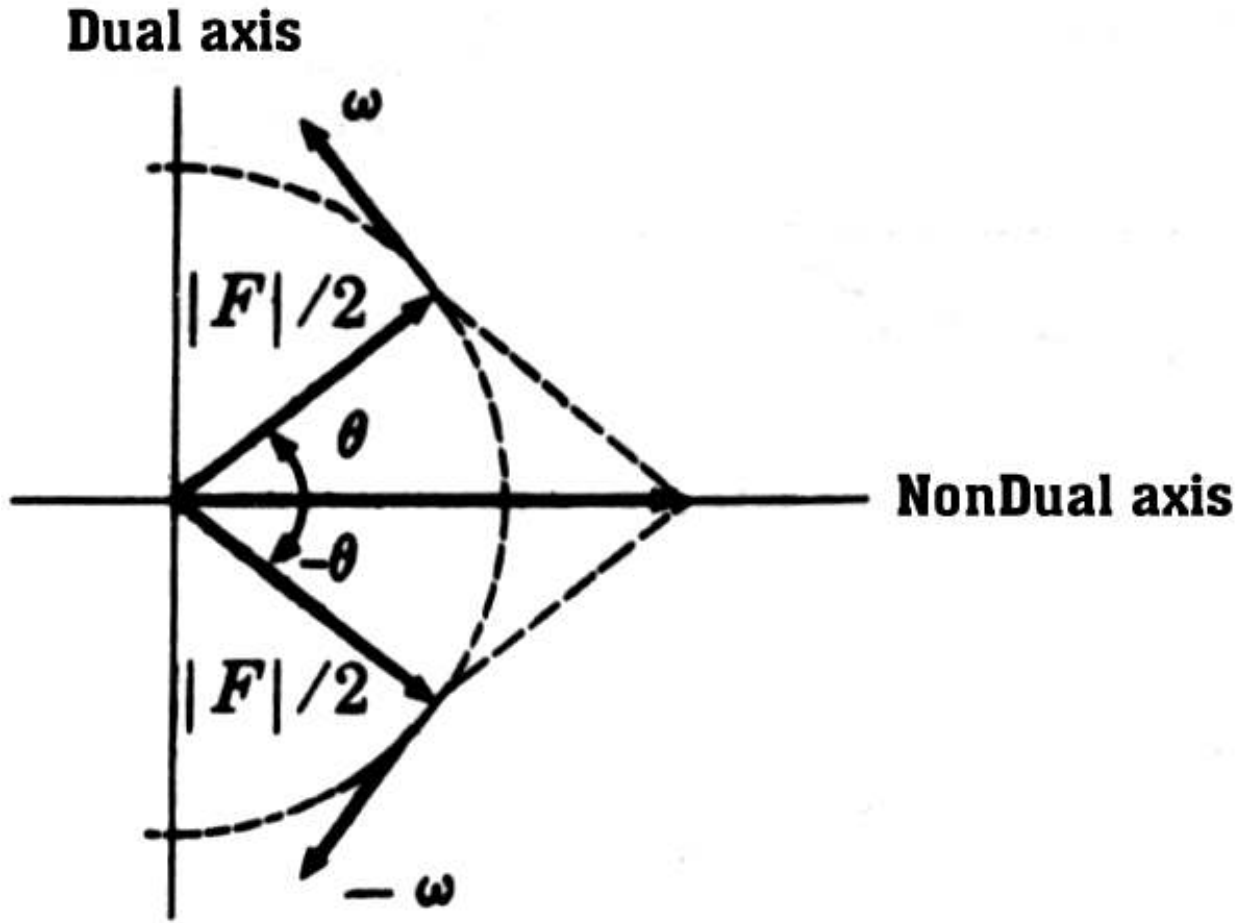

*Sine or dual nature of Euler Relation*

From that same Euler Relation we obtain the sine function represented by

$$\text{Sin}(wt) = [(e^{-J(wt)})/2 - (e^{J(wt)})/2] * J$$

so the sine function is expressed as the sum of two vectors rotating in opposite directions, one of them in counterclockwise or positive direction at an angular velocity w, and the second one in the clockwise or negative direction at an equal angular velocity w and as the vector rotate the two nondual components cancel each other, and as so the sum is a purely, as it were, "imaginary" or a dual vector, dual because by just changing the sign of the angle we can obtain the opposite.

The axis of the sine function is in fact the J axis(see how the sine function is affected by **J**), the so-called-imaginary-axis, or the symmetry axis, where symmetry is defined as similarity of forms or arrangement on either side of a dividing line, so on one side we have a positive magnitude and on the other we have a negative sign for that magnitude.

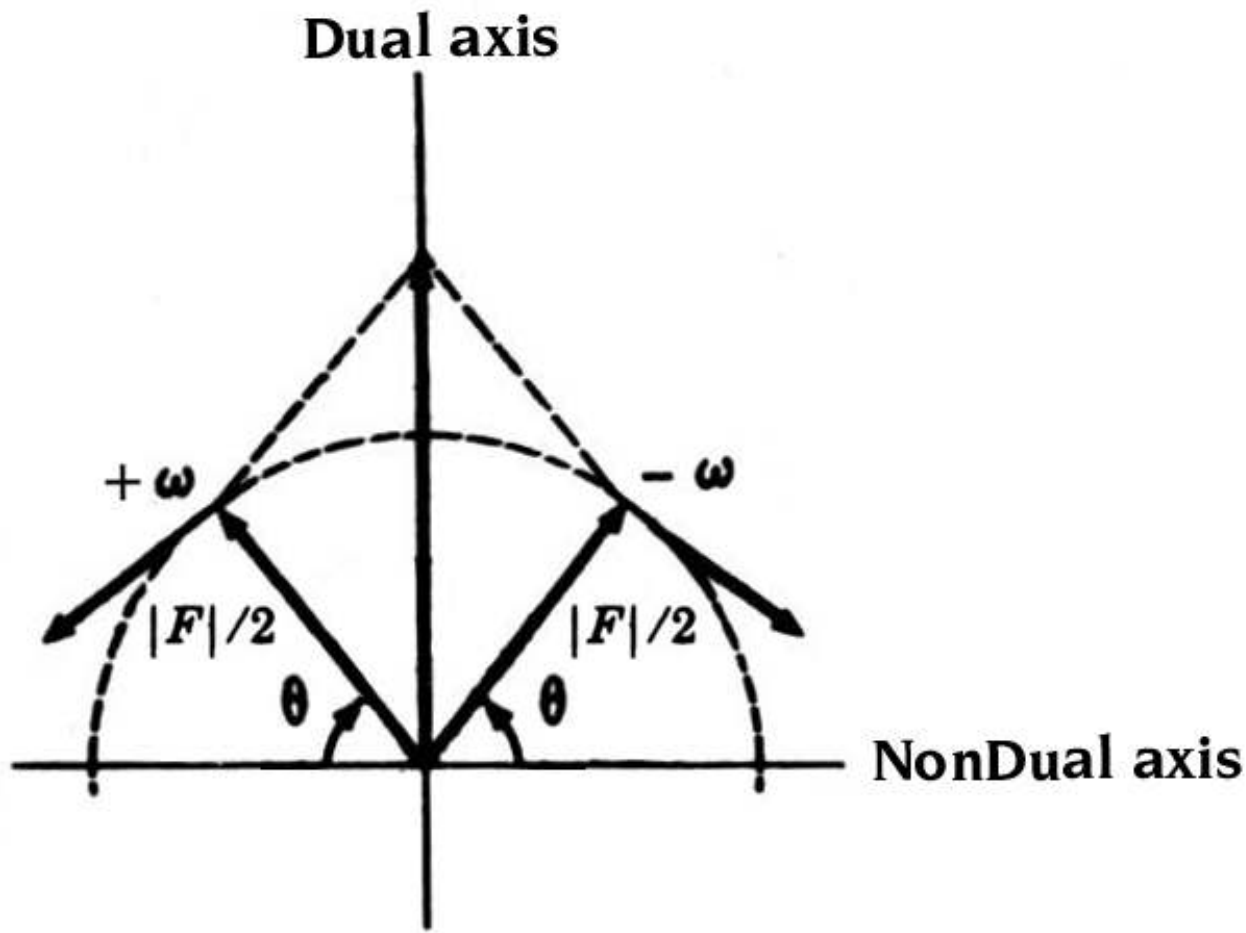

From this point of view it is not perfectly general to say that the sinusoidal wave, as an adjective, is the same for both sine and cosine functions even though we can obtain the one from the other by the addition of a phase angle of -90°; that addition is not at all a trivial one though. In fact to avoid any ambiguity we must differentiate clearly the asymetrical or nondual nature of the one, and the symetrical or dual nature of the other. Reality is in fact composed of two components, but in the physical domain, as it were, the " imaginary" domain, the dual nature is easily grasped and as so we can really say nature "loves" the sinusoid, and normally it hides as a mystery, its nondual nature. This nondual nature must always be discovered. It is some sort of fixed or nonchanging component to which the change of the system as a whole can be referred, but as a point of reference it must be discovered, it must be chosen or requires a decision. In this sense we can say that the foundation of all reality is an ultimate frame of reference in which the nondual, the nonchanging is at the background or up from the point of view of hierarchy and this implies then an "inclusive" attribute that is essential to have always in mind when dealing with reality. That ultimate frame of reference, we have named the domain of Form, permits us to define a domain independent of the observer and the object, as within it the object is embedded. We will then be able to have a Quantum Mechanics solution without the "observer drawback", as Karl R. Popper tried to find all his life but from the philosophical point of view and which was Einstein main concern about QM too.

Phase angle and magnitud are the two main state variables of that unit obtained with Euler relation and that has been named a phasor in EE. We will use interchangeably the terms vectors and phasors and in fact we will replace them by the Basic Unit System concept, *which in general will be a rotating entity in the complex plane at a given frequency*. In most practical problems all vectors are riding with the same frequency, so a sort of "merry-go-round" effect exists, in which vectors are seen as stationary with respect to each other, so the ordinary rules of vector geometry can be used to manipulate them as with that "merry-go-round" effect we obtain some sort of static framework within dynamism. But then we have also the chance to have the phenomenon of resonance. We are acquainted with such a phenomenon specially in the production of musical sounds and certain type of vibrations but the important point to recall is that *through resonance we can explain those cases where small changes can often produce large effects*.

We must recall additionally again the fact that the phase or the angle sign is associated with the positive or -negative sense of the vector rotation and as so:

- In the cosine case we can interchange both rotation vectors, by changing their sign and nothing changes, the form remains the same, they cannot be segregated just as in a magnet where we cannot differentiate the two polar components as they are seen always as one, as it were, the oneness property.

- In the sine case that interchange, or changing of signs of the angle, affects the final-resulting vector with respect to its position in relation with the nondual axis.

Resuming we have with ER a fundamental structure with two components separated by the radical **J**:

- one nondual in which oneness is an essential attribute and which brings us to mind that self-awareness capacity that permits us as humans to think about our own thinking process, that permits us reflectivity or to know that we know, as it were, a new way of "seeing"; a way to be conscious of our Being, that makes us different from the animal world. This nondual component is precisely that one that permits us to represent, as it were, the within of things. This within attribute was not possible to be considered in classical physics as it used to see reality just from the without of things. But with this within attribute we can envision a storing capacity of energy as that we find in magnetic fields, but also an information storing capacity in general. It is important to recall from the beginning that *this within attribute as a storing information capacity at its highest manifestation is very different from consciousness*, being more related with the fifth sphere of reality or just in plain mathematical words with the complex plane. So from here that pretension to reduce everything to consciousness is not our pretension anymore. A greater within means a greater complexity or a greater without, that can or cannot be necessarily a greater consciousness.

- one dual in which we can have two parts separated or clearly differentiated. Parts separated, is precisely that condition necessary for the application of analytical procedures, being the second one, the chance to linearize. We will see in what follows that ER most important isomorphic property is precisely *to reduce complexity to a minus one degree of complexity* making linear the representation of non-linear operators without reducing them.

But these two basic components are united in that mathematical symbolism called Euler Relation and at the same time separated in a radical way by **J,** repeating again at a higher level of representation the same basic structure we have found within that same relation, an inclusiveness attribute. We have then, as it were, a fundamental and basic minimum structural complexity represented by Euler Relation. Is this basic minimum structural complexity the one necessary to represent reality at its most profound structure? Does this basic minimum structural complexity give us those isomorphic properties needed for representing and deducing the most fundamental laws of physics and reality? The answer to these two questions will be yes, and this is what this proposal is all about.

# Complex Algebra and Isomorphic Properties

Vectors are ideal mathematical entities when relationships are important and as so we have vector sums and differences, but also multiplication, division, derivative and integrals and every one of these operations can be represented in the complex plane as some sort of complex metrics and even the resulting geometrical figures are simpler than those obtained in normal geometry. But the powerful advantage of complex number is seen in computations, or when using complex-algebra, where the isomorphic property of Euler Relation manifests all its co-variant power.

The main restriction we must pose from the very beginning is that of the same frequency, so that all rotating vectors considered must have the same angular velocity or frequency, so that we can have the "merry-go-round" effect. The frequency in this sense is that variable that can in fact produce large effects in a system that was previously chaotic before acquiring it. The order of this system depends on that acquired same frequency for all entities conforming the system. <span style="color:red">So this "restriction" is in fact the central point for the system acquiring *a higher ordered state*</span>. In fact that frequency is associated with one of the essential state variables of the system, the angle, being the other the magnitude. The state is in fact that one in which having a "merry-go-round" effect give us as a result, as it were, an organized complexity.

Rewritting ER again

$$e^{J(\emptyset)} = \cos(\emptyset) + J \sin(\emptyset)$$

we can note that it is a unit vector in which we have a length one, because

$$1 = \text{Sqr}[\cos^2(\emptyset) + \sin^2(\emptyset)]$$

So we have a unit vector standing at an angle Ø from the nondual or real or main axis

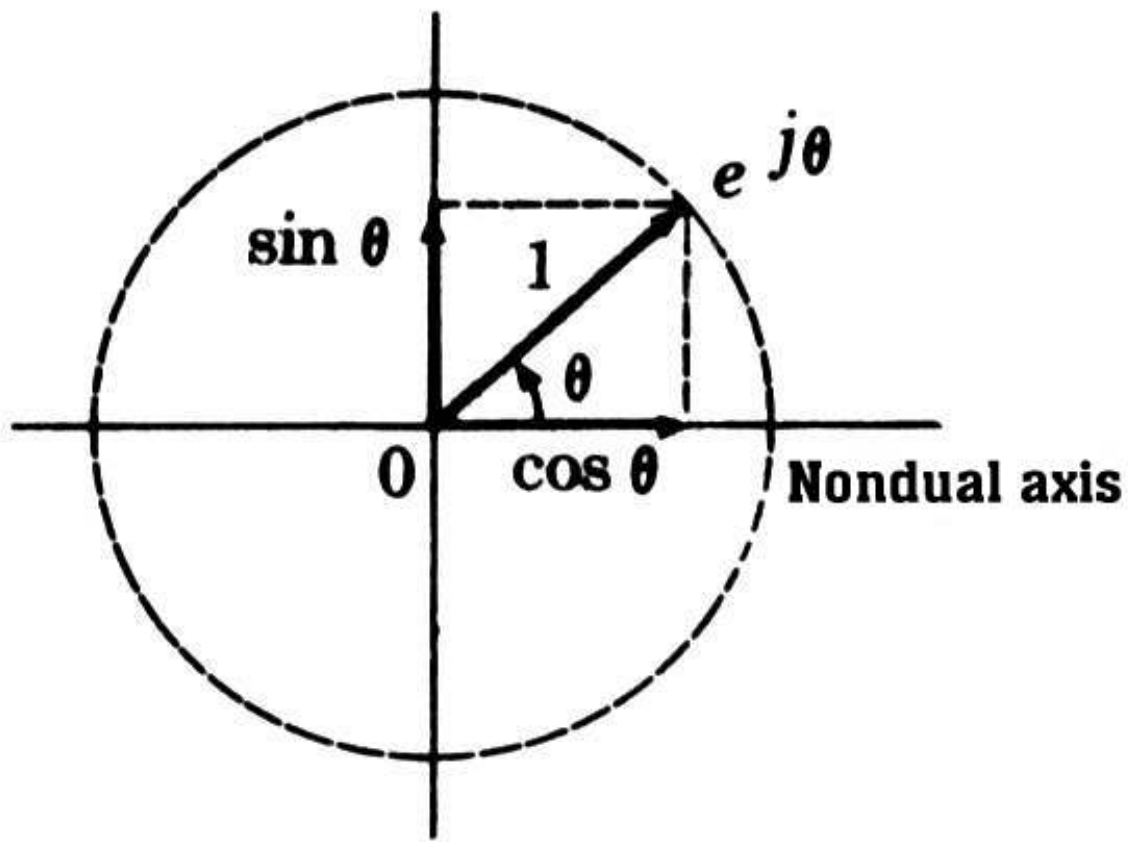

and we can have then in general, entities represented as

$$A = \text{Abs}(A) * e^{J(\emptyset)}$$

that can be represented too in a rectangular form taken from the figure

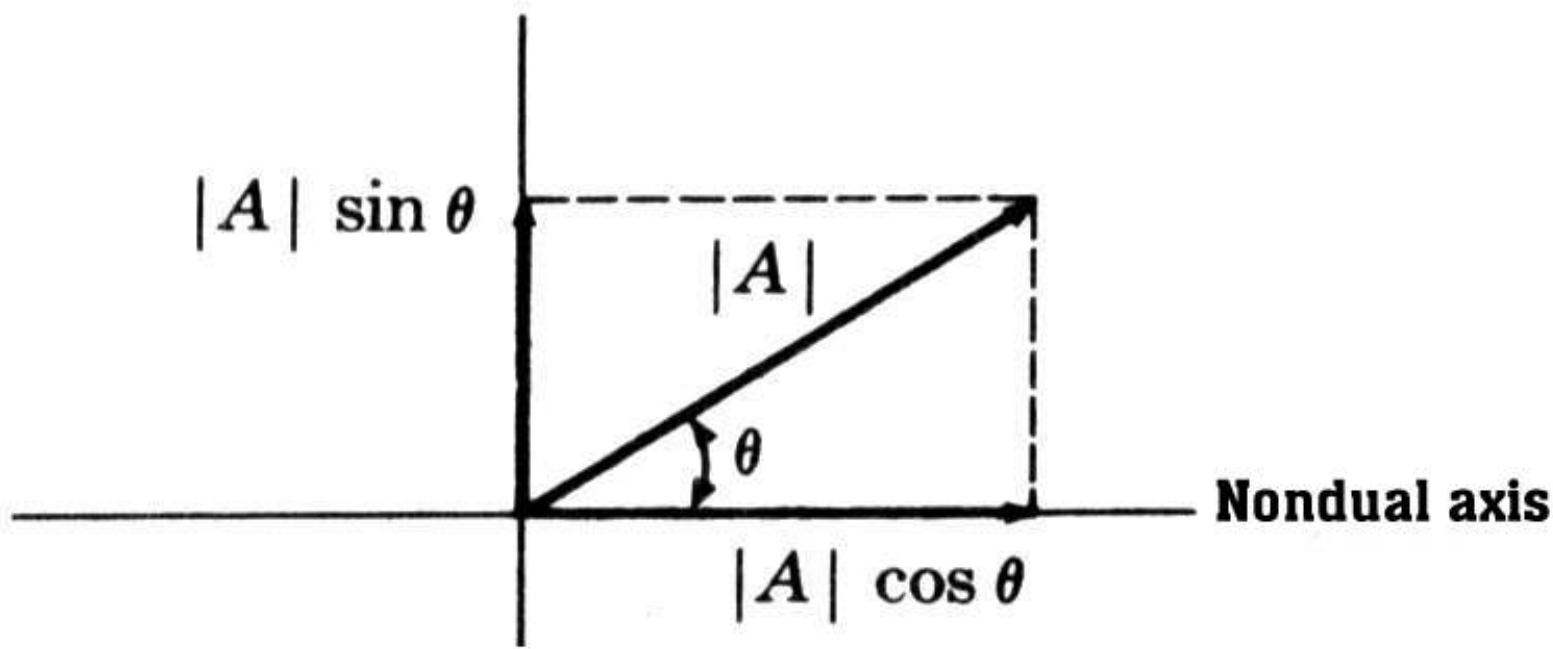

as

**A** = a-nondual + **J** * a-dual

where

A-nondual = Abs(**A**)*Cos(Ø)

and

A-dual = Abs(**A**)*Sin(Ø)

With these expressions the complex algebra can be established having in mind that we must not mix, as it were, oranges and apples, we must in fact abandon once and for all the reductionistic tendency, with this complex representation. We must make in fact a radical distinction between the two basic components that are used to represent reality and according to this radical duality, the nondual components must be used with nonduals, and the dual components with the dual components, noting that

**J** * **J** = -1

in case of multiplication and division of complex vector entities.

With this in mind, all laws of normal algebra and arithmetic are preserved as with this radical separation we preserve homogenety on the one side and heterogenety on the other, and as so we then have another rules for the basic operations with complex algebra. We must differentiate the dual nature of reality to integrate them again, through the basic unit system concept, that transforms itself in a powerful explaining and simplier tool, some sort of a truly isomorphic unit.

***The sum or difference of two Buses***. To obtain the sum of two Buses, we must decompose it in its two basic components in rectangular form as:

**A** = a-nondual + **J** * a-dual

**B** = b-nondual + **J** * b-dual

so

**C** = **A** + **B**

equals

**C** = ( a-nondual + b-nondual ) + **J***( a-dual + b-dual )

so

c-nondual = a-nondual + b-nondual

and

c-dual = a-dual + b-dual

in the same way for the difference

***The product of two Buses.*** To multiply two Buses we must use instead of the rectangular form, the corresponding polar form as:

$$A1 = Abs(A1) * e^{J(Ø)}$$

$$B1 = Abs(B1) * e^{J(Ø)}$$

so we have

$$A1 * B1 = Abs(A1) * Abs(B1) * e^{J(Ø1)} * e^{J(Ø2)}$$

or

$$A1 * B1 = Abs(A1) * Abs(B1) * e^{J(Ø1+Ø2)}$$

a resulting vector **C,** in which their magnitudes are multiplied, as in normal arithmetical cases, but their angles are summed, so a multiplication or a nonlinear operation is transformed, as it were, in a linear operation.

$$C1 = Abs(C1) * e^{J(Ø3)}$$

where

$$Abs(C1) = Abs(A1) * Abs(B1)$$

and

$$Ø3 = Ø1 + Ø2$$

By similar mathematical reasoning we can obtain the quotient of two buses, in which that quotient transforms itself both in a difference of the corresponding angles and a quotient of its magnitudes. The magnitudes behave themselves in the same way normal mathematical arithmetical quantitative operations behave. All its rules are preserved.

But if we call qualitative those aspects related with the phase angle, we see then, they are summed instead of multiplied in case of complex algebra representation. This capacity of complex numbers to reduce a nonlinear operation to a linear one, is in fact the one that makes them such a powerful simplifier tool, and adequate to represent the complex nature of reality we have been presenting from the beginning of this paper. But it is important to recall that linearization does not affect at all the quantitative part, just the qualitative one, giving us *a new capacity to think the whole correctly* and the chance to introduce bodly in our intellectual frameworks new categories, so a new world outlook is obtained that includes in itself an explanation and development of new things and even solves once and for all the paradoxes of wave and particle as we will see when deducing the complex Schrödinger Wave Equation, but also when obtaining the exact mathematical representation of the most intriguing open system mankind has ever known and which had always exerted a fascination for him, I mean, the pendulum movement.

*Conjugate of two complex numbers.* Two buses are conjugate if they have the same magnitud and equal angles, but each angle being the negative of the other. As we have pointed out that negative sign has to do with an inversion of the rotational sense in such cases where the angle is a function of time.

*Differentiation and integration.* We have already pointed out that remarkable property of ER we have qualified as isomorphic and that has to with the fact that the integration and differentiation of ER, are themselves ER of the same frequency. And in that case in which the angle of ER is not complex, but just a simple function of a variable, the differentiation and integration of that ER function is the same, in both cases. There is no way to distinguish between them. But that differentiation appears clearly in those case we have a complete ER.

By the rules of elementary Calculus, the derivative of an ER with a constant magnitud, has the same magnitud as the original vector multiplied by the frequency w, but the angle has been advanced by 90°, or just the expression multiplied by **J,** which is an ER, where the angle is 90°.

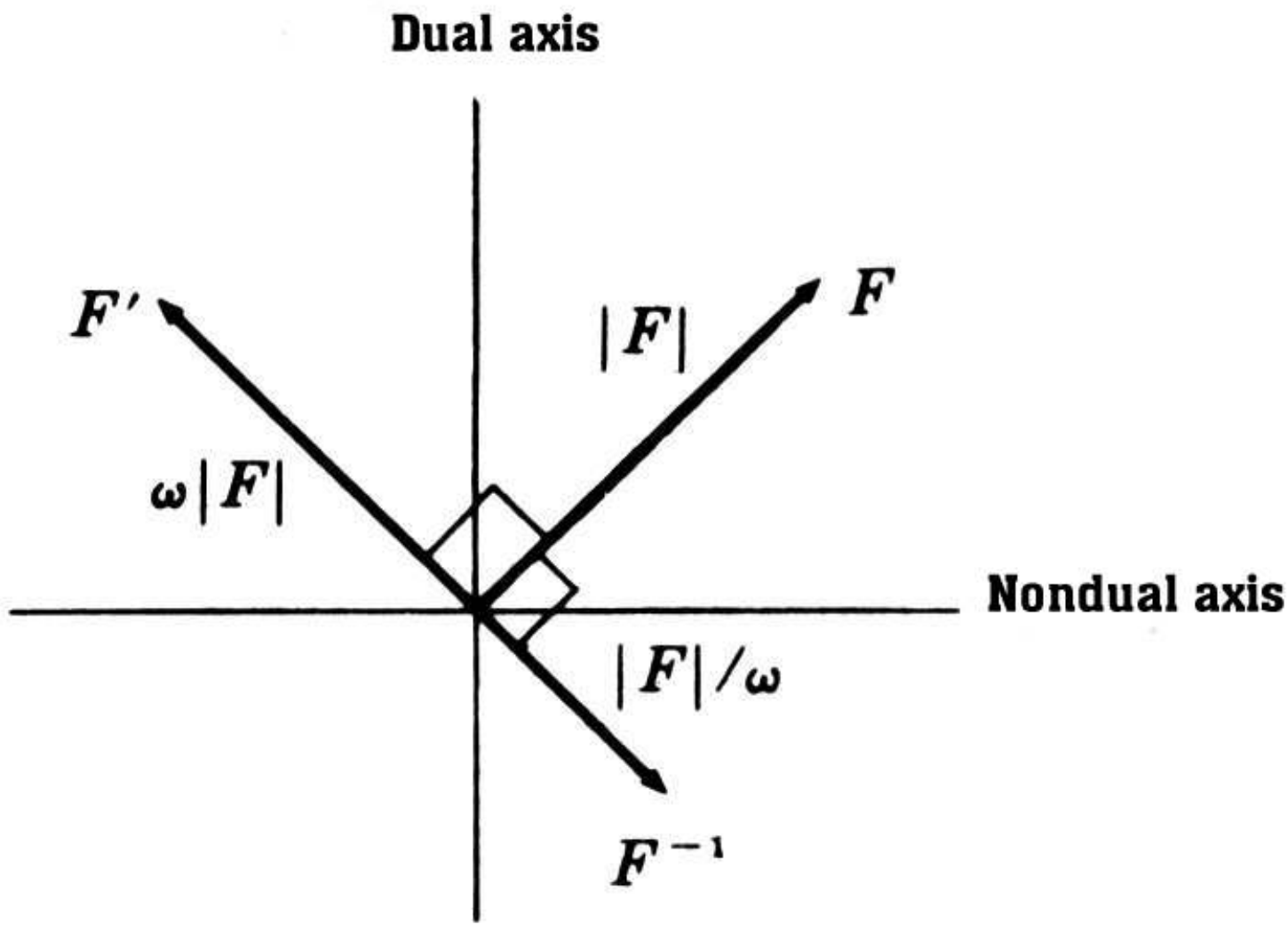

And the integration is the dual operation as that of differentiation and as so the integration of an ER with constant magnitud, has the same magnitud as the original vector divided by the frequency w

**F $^{-1}$** (INTEGRATION)

**F '** (DIFFERENTIATION)

but the angle has been retarded by 90°, or just the expression multiplied by **-J**, as were, an opposite sense of rotation. This way of reasoning is exactly the application of what I like to name the duality principle, but which is a way of applying a binary logic or symmetry to obtain a dual reality from one of its components.

The complex vector algebra simplifies mathematical operations by one degree, so in this way it reduces complexity by one degree too and as so it becomes a real isomorphic tool as even complex operations such as differentiation and integration are reduced to simple algebraic operators. This technique of reducing complexity by minus one degree is the most important characteristic of the Electrical Engineering mathematics, where phasors are its "bread and butter" tool. This reduction of complexity by minus one degree permits us to comply with those two conditions of analytical procedures:

- linearization and

- to have weak interaction between the parts in those cases in which the general open system is closed,

which was what Ludwig Von Bertalanffy pointed out as the central and methodological problem of systems theory.

But the important thing to recall is that Euler Relation is the ideal isomorphic unit, as the sums, differences, integrals, and derivatives of Euler Relation functions of a given frequency are themselves Euler Relation functions of the same frequency, they do not change their form with these operations, they remain invariant or just co-variant. **No other mathematical function is preserved in this fashion**, as so it is the ideal one for filling that requirement Einstein put in his **The foundation of the General Theory of Relativity**, 1916, when he wrote "The general laws of nature are to be expressed by equations which hold good for all systems of co-ordinates, that is, are co-variant with respect to any substitutions whatever(generally co-variant)".

We can then define a differential complex geometry in which we have a differential of reality defined as a Basic Unit System or a Holon such that

$$DS = Abs\,(DS) * e^{J(\emptyset)}$$

with this complex metrics we can find and deduced all the fundamental equations of physics. It is not strange to find such a powerful explaining tool after we have found all those isomorphic properties of complex numbers. A Holon has embedded a within and a without, as it were, an analytical and a synthetic capacity, or a partness and a wholeness attributes. Leonard Euler must have felt a special feeling of wholeness when he found this relation.

## The Principle of Synergy

The late Abraham Maslow was the one who coined the term "synergy", an obscure term from anthropology, but he used it for the first time in business to describe how wealth can be created from cooperation. Creation of wealth, emergence of new things and structures that modify themselves to give better performance are main issues of the so-called-sciences-of-complexity or systems sciences. Structure and environment or as we would say, threeness(Oneness) and openness, they both claim for a whole that is greater than the sum of the parts, or for a basic and fundamental framework in which three fundamental and interdependent entities are put in mutual interaction. They pose the need to have as starting point not only a minimum structure, but also a minimum system of elements in interaction.

In the physical domain synergy can be found whenever we have a real transformation of energy to another useful way or level which presuposses interchange or transformation of energy between systems. This "useful way" is in fact the real source of new applications and in this sense synergy is there, where we have emergence of new realities or just open systems that have exchanging-energy-capacity with the environment. That whole greater than the sum of its parts has nothing to do with metaphysic, as that "greater" comes from the very nature of an open system.

Electrical energy, or the AC we use at home, is in fact the result of the application of the Principle of Synergy where energy is transformed from a primary source, hidraulic for example, to electrical energy by moving a threefold magnetic structure, which gives at the output the AC energy we can utilize in many ways. In these cases the open system concept is concomitant.

If we consider three-double-complex-vectors or phasors, or just Buses constituted each one of them by

$$A = Abs(A) * e^{J(\phi)} + Abs(A) * e^{J(wt)}$$

the first vector

$$A = Abs(A) * e^{j(\phi+120)} + Abs(A) * e^{j(wt)}$$

the second vector

$$A = Abs(A) * e^{j(\phi+240)} + Abs(A) * e^{j(wt)}$$

the third vector

where they correspond to three physical counterpart dispositions displaced 120° from one another just like the figure and where that physical disposition cancels, as it were, its apparent cause so that the cause-effect relationship is subtle or the cause become hidden, or just it depends on the interchanging of energy or information with the environment or field, so at last we have

$$A' = Abs(A') * e^{J(wt)}$$

a rotating phasor or Bus at a given frequency.

*A given frequency* is precisely the one attribute that makes it possible, the "merry-go-around" effect, as it were, that new order or reality that comes from the application of the Principle of Synergy. If the frequencies are not the same for the three independent entities we wont't have the desired "merry-go-around" effect; it is just obtained when that frequency is reached, some sort of emergence state, in which one small effect -a small change in frequency- gives us a large effect or a new emergent order.

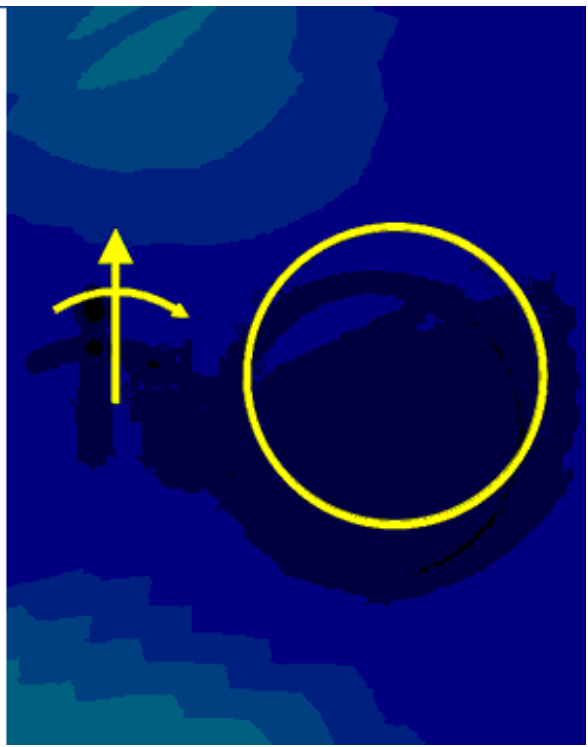

But those three or six independent entities in interrelationship conform just one entity or a whole picture, or a form represented by a circle.

We have already come across with a radical separation of reality, in which we have a without and a within but embedded in a whole, a whole that is not just the result of an arithmetical sum of elements. The Principle of Synergy or that glueing principle can be just represented by a fundamental sevenfold structure as the one of the figure in which we have not only a theoretical framework that has embedded that radical separation, but also a practical way of representing that minimum complex structure of nondual and dual, and in which we have a new type, as it were, of constitutive characteristic whereby we must not know only the parts, but also the relations.

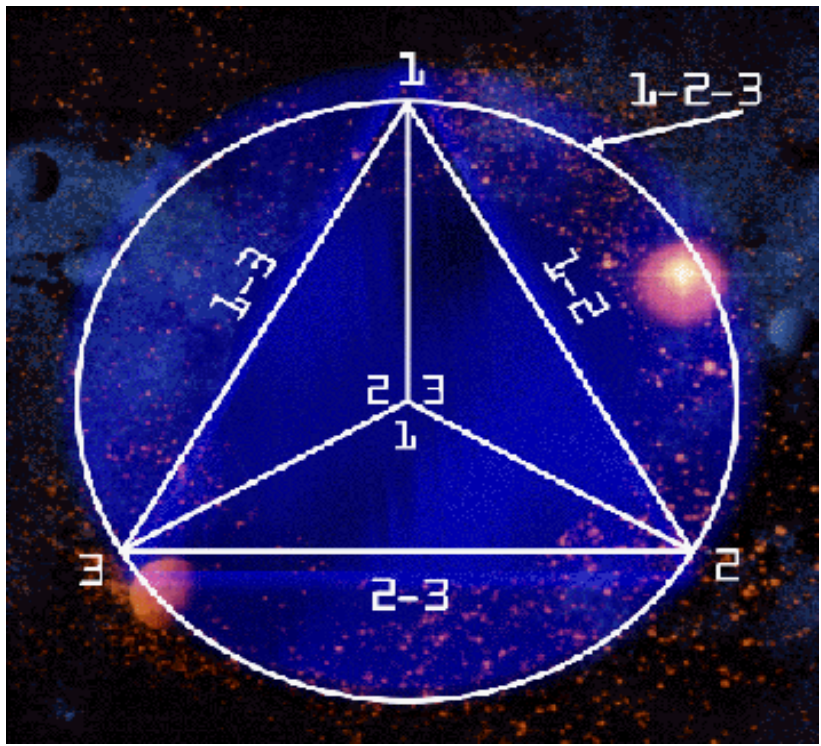

But the point to recall is that structure represented by

- three inner or nondual relations, or just the 1-1, 2-2 and 3-3 relations

- three outer or dual relations, or just the 1-2, 2-3 and 1-3 relations

But additionally the wholeness attribute is associated as we have pointed out with an emergent property, represented by the 1-2-3 relation. But here we have a clear departure of a holistic framework, as even though we have that holistic property embedded too in this framework, that important attribute is embedded in a Basic Unit System or Holon, not in a general abstraction that is just the whole, and where the hierarchy-relationship between that part and the whole is not cleat at all.

The Bus(Holon) concept is embedded in a formal complex mathematical framework, whose generality is associated with those isomorphic properties we have already presented and that give us a real internal consistency which allows to treat steady-state systems by the same general techniques or methodology. Furthermore those six relations are some sort of detailed complexity as opposed to the 1-2-3 relation which is the complexity of the whole. The traditional problem between the part or detail and the whole is not a problem anymore with this sevenfold structure that can be used at different levels of reality without reducing the one to the other.

The fact we use a complex mathematical symbolism, as a tool, for representing this sevenfold structure avoids that syncretic whole reasoning Galileo used when he wrote *There are seven windows given to animals in the domicile of the head, through which the air is admitted to the tabernacle of the body, to enlighten, to warm and to nourish it. What are these parts of the microcosmos? Two nostrils, two eyes, two ears, and a mouth. So in the heavens, as in macrocosmos, there are two favorable starts, two unpropitious, two luminaries, and Mercury undecided and indifferent. From this and many other similarities in nature, such as seven metals, ect., which is were tedious to enumerate, we gather that the numbers of planets is necessarely seven.* But it also prevents using as a key guiding principle *the primacy of the whole* which results in arrogant role of dominance when applied to organizations.

## The Pendulum

The pendulum movement is so remarkable not just for its role as a cornerstone in the birth of modern science that permitted Galileo a paradigm shift, but also because it is the most natural example of an open system. When looking at the swinging body he saw a body that almost succeded in repeating the same motion over and over ad infinitum. Its succes in repeating the same motion over an over lies in its interrelationship with the environment or in its openness attribute. By looking at the pendulum *Galileo reported that the pendulum's period was independent of amplitud for amplitudes as great as 90°, his view of the pendulum led him to see far more regularity than we can now discover there.* Is this view of seeing more regulates than there really existed part of the incapacity of normal science to explain pendulum-like regularities in a most documented way, and following a mathematical methodology with no leaps? Or *How else are we to account for Galileo's discovery that the bob's period is entirely independent of amplitude, a discovery that the normal science stemming from Galileo had to eradicate and that we are quite unable to document today.*

The exact simple pendulum solution implies the solution of a first order differential equation which implies too an integration whose solution is an elliptic integral. This means the introduction of an approximation factor that could only be found by observations of the pendulum real behavior, some sort of trial and error procedure. In fact normal mathematical symbolism, I mean, not-complex-mathematical symbolism, cannot give reason of that approximation factor without using some sort of methodological leaps, to explain deviations. Normal science works with closed systems, that is, systems that do not exchange with their environment. But the pendulum movement seems to violate even that infamous second law of thermodynamics. In fact it gives us a natural sense of eternity just as the poet *Jacques Bridaine* wrote

*"Eternity is a pendulum whose balance wheel says unceasingly only the two words, in the silence of a tomb, 'always! never! always!...'"*

After exploiting the cyclical wave nature of Euler Relation in EE, it is obvious to expect we will be able to explain with it, all those natural phenomena such as that of the pendulum, in which we have cyclical or wave movements. But for achieving this, it is necessary to realize a real paradigm shift. With the complex plane we have introduced in fact a new sphere of reality in which we have embedded both the nondual and dual nature of reality, and as so a complex metrics, whose main characteristics are:

- on the one side, its isomorphic property, as we have extensively pointed out up to this point, and which gives us a powerful methodological tool to explain those cases in which real dynamism is involved

- but also that property we have associated with a "merry-go-round" effect, and that permits us to make a mathematical representation of real dynamic entities or of the generation of forms, as is the case with the pendulum, in which its form is continually generating itself just as a result of a mere impulse.

A new sphere of reality or the sphere of form, in which the four-dimensional space-time continuum is embedded. A new category or totality that contains the physiosphere is also that sphere in which we can have animated forms, as it were, the biosphere, or the Bergson-Theilhard de Chardin DURATION concept.

A complex metrics is then a top-down metrics and this also means an axiomatic nature, so from the notion of the Basic Unit System and those normal regulaties already known from physics, but also from those properties obtained from geometry we can obtain the state of our Bus system, through a methodology in which instead of starting with the postulation of a differential equation, we start with the Bus concept, as a tool to integrate or obtain that state. And this will be our aim for the pendulum case.

From the point of view of the BUS, the pendulum movement is a rotational motion. The pendulum as a Bus is then an open steady state system and the earth gravitational field is its context, and when observing that cyclical movement we can observe a maximum angle Ø, or Ømax, for a corresponding maximum displacement Smax. Solving this problem means, integrating , some sort of constitutive characteristic, not a summative one.

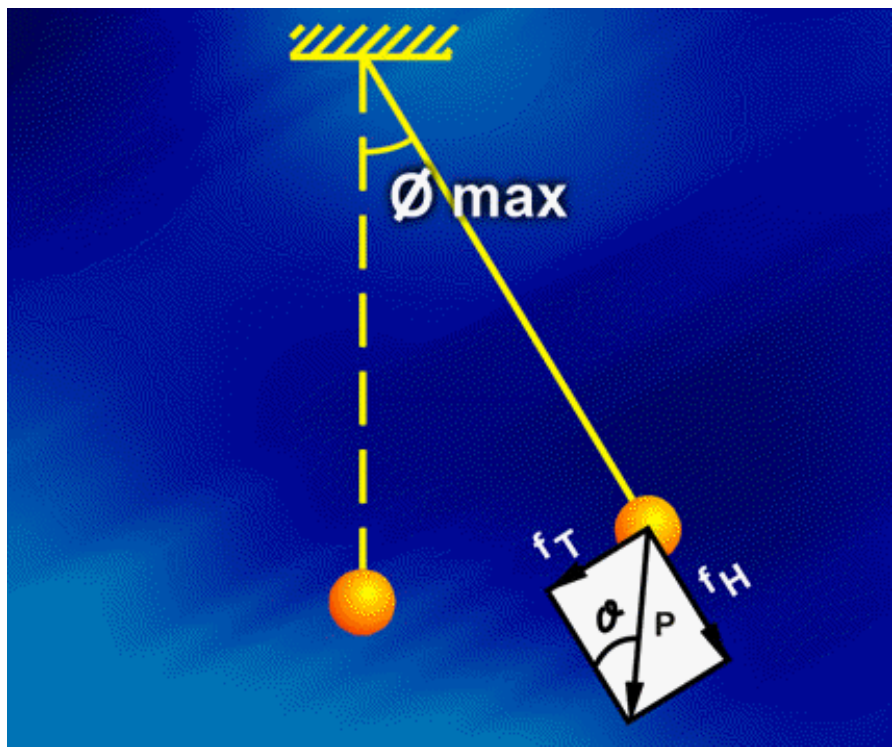

We will use Int [ ] as the integration symbol - so the complex trajectory DS between

$S = 0$

to

$S = Smax$

will be

$$\text{Int}_0^{Smax} [DS] = \cos(\emptyset max) * \text{Int}_0^{Smax} [DS] + J * \text{Sen}(\emptyset max) * \text{Int}_0^{Smax} [DS]$$

so we have

$$S = Smax * e^{J(\emptyset max)}$$

but additionally we note that this is a static and potential expression related with space, and as so we must introduce its dynamic counterpart by multiplying both part of the equation by the basic unit of time given by:

$$e^{J(wt)}$$

where

$\emptyset = wt$

So reality S of the pendulum is generated and represented by:

$$S(t) = S_{max} * e^{J(wt + \emptyset max)}$$

Intuitively, from geometry we know that in a circle we have

$$S_{max} = L * \emptyset max,$$

where L is the radius or the cord length of our bob system and

$\emptyset max$, that angle subtended by the arc, so we have

$$S(t) = L * \emptyset max * e^{J(wt + \emptyset max)}$$

as a general dynamic expresion in which the principle of synergy is included. Its manifestation can be recognized in that rotating expresion.

## Balancing Equality and the Pendulum

To obtain the pendulum harmonic motion or just a steady state we must apply a balancing or a compensator equality, between two forces or polatities:

- an inner one related with the weighting mass of the bob and

- an outer one related with the so-called-inertial mass.

This equality played a very important role in Einstein works about gravitational fields. Up to that time that equality had been considered a mystery in the Newtonian framework. It generated those famous mental examples with elevators, in which bodies are subjected on the one side to an inertial force, and on the other to the gravitational field.

When these two forces or polarities within a comprehensive whole become equal then we have a harmonic motion or a motion that almost succeeded in repeating itself over and over ad infinitum. A balancing loop which implies a rotational motion or the Synergistic Principle means that the law of opposites must always be considered in a comprehensive whole, or dynamic context, producing a cyclical movement. But if we consider inertial forces then we must apply Newton Second Law, but this time at the fivefold continuum, or complex plane.

The first derivative of S(t)

$$S(t) = L * e^{J(wt + \emptyset max)}$$

is

$$dS/dt = L * \emptyset max * e^{J(wt + \emptyset max)}$$

and the second derivative:

$$d^2S/dt^2 = -(L \cdot \emptyset_{max} \cdot w \cdot w) \cdot e^{J(wt + \emptyset_{max})}$$

being amax = ( L * Ømax * w *w *) the maximum absolute value of acceleration.

According to that second law we have at the maximum point Smax the inertia "vector", which tends to maintain the motion as:

**F**max = m *( L * Ømax * w * w)

if we apply, on the other hand, the vector composition of forces and using tangential components at that point we have the mass "vector" as:

**F**max = m * g * Sen(Ømax)

so equating both the inertial and the mass "vectors" we've got:

L*Ømax*w*w = g*Sen(Ømax)

so the angular velocity is:

w = sqr( g * Sen(Ømax) /( Ømax * L))

but if

w = 2*Pi *f

and

T = 1/f, where f is the frecuency and T the period then

T = 2*Pi *sqr((L/g) *( Ømax /Sen(Ømax))

which is the exact pendulum formula.

The factor sqr( Ømax /Sen( Ømax) ) tends to one when Ømax is small and it can be omitted as long as the amplitud does not exceed **10°**. In Table 1, in the second row we can see the values K of the elliptic integral, in the third row the corresponding factor according that integral, and in the fourth row the corresponding factor obtained with the Bus concept. In the fifth row we can see the difference in error between the elliptic factor and the Bus Factor. This difference in errors as the amplitud increases can be taken as a fallibility criterion for those angles greater then 30 degrees where the error between the two is definitely inaceptable.

Table 1. **Aproximation Factor for the Period of a simple Pendulum**

| Ø(angle) | 0° | 10° | 20° | 30° | 60° | 90° | 120° | 150° | 180° |
|---|---|---|---|---|---|---|---|---|---|
| K | 1.571 | 1.574 | 1.583 | 1.598 | **1.686** | 1.854 | 2.157 | 2.768 | Infinite |
| 2K/Pi | 1.000 | 1.002 | 1.008 | 1.017 | 1.073 | 1.180 | 1.373 | 1.762 | Infinite |
| Ø /Sen(Ø) | 1.000 | 1.005 | 1.002 | 1.047 | 1.209 | 1.570 | 2.418 | 5.235 | Infinite |
| Error(%) | 0.000 | 0.298 | -0.595 | 2.865 | 11.248 | 24.84 | 43.217 | 66.34 | N/A |

The exact solution introduces a nonlinear character in the pendulum motion which gives reason of its real behavior defined by the well known observed laws. Normally this problem was solved by resolving a differential equation, that gave finally an elliptic integral, or some kind of solution that must be validated with the real observed behavior of the pendulum, as an elliptic integral cannot be expressed in terms of the usual algebraic or trigonometric functions, see [Vector Mechanics for Engineers](). Galileo did not know this factor, and so he extended the application of the observed pendulum laws far beyond its real range, even to 90 degrees. But the fact that the time of any amplitude were independent of the mass of the body made he think in the falling of bodies and specially in the well-known Pisa Tower experiment, in which both an iron and a wood sphere fall with the same time, which means in all cases, the trajectory followed by those bodies is an invariant.

The pendulum is essentially a device for measuring time. It is in fact, as it were, the contrivance of time. Its form unfolds itself in time, it is the dynamic device for excellence. Differently from those movements of classical Newtonian mechanics, like those of the planets, in which time does noy play really any role and in which we could talk about the chance to invert time, the pendulum movement in this sense is not a classical movement, and it answers the question posed by Thomas S. Khun about that movement. So the Principle of Synergy applied to the pendulum explains why this principle seems opposed to that infamous second law, as with it, forms can be generated and we can obtain in natural way a steady state open system.

## Quantum Mechanics

The Quantum Mechanics problem before being a scientific problem is a philosophical problem as from the beginning it touches the same nature of reality and the relation of the subject with that reality. The relation between object and subject, the active factor of the subject in the process of cognition is in fact an epistemological problem that has to do with the way the subject defines what he understands by objects. But this philosophical problem can be found in the idea of the active role of language in the shaping of our worldviews or images of the world, and as so it is primitive and was the main concern not only of philosophers but also of historians as we can see in Adam Shaff work History and Truth. And we know also the efforts done by Karl Poper during all his life and specially in the introduction of his [*Quantum theory and the Schism in physics. From the Postscript to the Logic of Scientific Discovery*](), to exhorcise consciousness or the observer from physics.

Reality is independent of human mind, of the subject perceiving it, and we can distinguish in it three great realms:

- Reality per se

- The subject that perceives that reality

- the object perceived by the subject as reality

The structure of reality perceived by the subject is crucial in the determination of the objects of study, that as a matter of fact, are those objects normally studied by science. But science is a product of human activity and as so language is crucial in the determination of those objects of study. This problem took Karl R. Popper to define a third world, as an intermediate world between consciousness and objective reality.

As we have seen we can have two kinds of mathematical languages, a partial one that not necessarily denies complex numbers as it uses them in "convinient ways", and an integral one in which complex number are used for building a unit in which the radical J is some sort of operator to distinguish between two different orders of reality, as it were, the nondual and the dual nature of reality embedded in that same unit. In this sense this is not a new theory but a way of "seeing" and interpreting reality in which uncertainty is always concomitant.

When we talk about Reality per se we are not meaning a third world as that of Plato, divine, superhuman and eternal, but of reality as defined by Mortimer J. Adler in his **Adler's Philosophical Dictionary** where he wrote:

*BEING The word "being" is an understanding of that which in the twentieth century is identified with reality.*

*What does the word "real" mean? The sphere of the real is defined as the sphere of existence that is totally independent of the human mind...Another distinction with which we must deal is that between being and becoming, between the mutable being of all things subject to change, and the immutable being of that which is timeless and unchangeable. That is eternal which is beyond time and change. In the real of change and time, past events exist only as objects remembered, and future events exits only as objects imagined.*

Historically the prevailing structure of reality has been a dualistic one in which reality as a whole is divided in just two domains, being each one, as it were, parallel to the other, with no possibility of integration in case of gross dualism. Normally there has been some sort of partialness depending on the accent put either in philosophy as the queens of all science, or in science as the real-objective science, or just two and only two realms of reality. Philosophy and Ontology are not considered normally as independent domains, but for our purpose and aim, philosophy must look for an integrated image of reality and in the searching of methods of generalization so that philosophical theses are kept from contradicting science, in other words it must contribute to oneness and not to partialness. On the other hand Ontology must look for the more-being, and as such realism must be concomitant as a main issue.

In our context, in which we use a structure defined by a basic unit system, those objects of study can be classified as real open systems and as ideal closed systems. An open system can have though a variable center and as such its external manifestation or its form cannot be determined, as its field or circle of influence is variable, and this will be the case for the electron. Other open systems in which both its center and its radius are not variable will have a more determined state and will be near to an ideal closed system. But this issue has to do directly with the Uncertainty Principle, in which we have as in philosophy no univocal conditions as those we can obtained with closed systems. But the important point is the correlation we have between closed systems, measuring and open system and the Uncertainty Principle. The incapacity to measure the momentum and the position of an electron at the same time, has been associated with a philosophical problem that has to do with an emphasis on the subject doing the experiment, pointing out the fact that the structures of the subject are crucial in the way he defines reality or its object of study, which is similar to that philosophical problem in which language is considered as a factor that creates reality.

We can talk about metaphysics when we have assertions for which we don´t have procedures on the one side, or on the laws on the other, that can be used to verify those assertions. In the latter case when we have laws we can built measurement instruments to verify the Data. On the other hand we have known procedures that when used they take us to a series of consistent results, that can be described qualitatively, but not necessarily quantitatively. In both cases our main task as scientists and philosopher must be the searching of thruth.

In our case we will consider this problem of the Uncertainty Principle, in case of an electron, as the impossibility to determine the state of the corresponding Basic Unit System because there is not a law with which we can interrelate its two components, as it were, its dynamic counterpart and its static one. In this way we can separate the scientific problem of Quantum Mechanics from a philosophical problem which I think was the aim of Karl Popper when defining his "third world theory". In our case there are clear differentiation between precision and uncertainty, as there are cases such as the electron case in which there is no way to reduce the qualitative to the quantitative, which was Karl R. Popper requisite to increase the degree of contrastability of certain theories.

The trajectory or more appropriately the reality of an electromagnetic entity as that of the electron, which as a matter of fact behaves as a wave, can be represented by the bus concept as

$$DS = Abs(DS) * e^{J(\emptyset)}$$

But a general conceptualization of the Bus concept can be seen as a mathematical expression of energy in its most primary definition, but energy "is like a frequency multiplied by Plank's constant h"

$$E = h*f$$

and that angle ø can be replaced by

$$\emptyset = 2*p / h * (p* x - E *t)$$

a well-known expresion taken from Quantum Mechanics

and

$$1/ \lambda = p/h$$

where $\lambda$ is de Broglie wave length, h is Plank constant, and p is the momentum of the particle - a Bus rotating at a very slow unknown rotational speed in the complex plane- that can be also written as

$$p^2 = 2*E*m$$

The two state variables of the BUS in this case are x and t, so to able to determine its state, we must find or know a relationship between them. The geometric-like behavior of the pendulum gave us the clue to find that relationship. In case of the planet movements it is the second Kepler Law associated with a central force movement the one that permits to determine its state.

We know that for the electron we cannot find a relationship between its two state variables, and this is another way to present the Uncertainty Principle which means from the geometrical point of view we will not have a known trajectory followed by the electron, as was the case with the pendulum or as it is with the planets. In both of these cases we have a "real" differential equation associated with. In the electron case we have the well-known complex Schrödinger wave Equation that was presented by him in 1926 as a postulate, that is, as a way of saying that equation does not have just one solution starting from initial conditions, so the need to measure was replaced by qualitative methods, where one must focus in a behavioral area and not in finding laws that permit us to measure. In case of the pendulum its form changes with time, but at the same time, its centerness attribute is localized, which seems not to be the case for the electron case.

But our aim is to deduce that complex wave Equation in the context of the Bus concept. Let us suppose an unknown general solution, as a function of space and time, as it were, of its two state variables

$$S(x, t) = Abs(S) * e^{J(2\lambda/h * (p*x - E*t))}$$

let us rename

## Abs(S) by $\Psi$

and

## S(x, t) by $\Psi(x, t)$

to distinguish magnitude from complex quantities on the one side, and on the other to use the traditional notation used when representing Schrödinger Wave Equation in Quantum Mechanics.

Expressing E as function of momentum and replacing

$E = p^2/2m$

we have $S(x, t)$ as

$$\Psi(x, t) = \Psi^* e^{J(2p/h * (p*x - (p^2/2m)*t))}$$

$$\Psi(x, t) = \Psi^* e^{2pJ*p/h*x} * e^{-2pJ/h*(p^2/2m)*t} \qquad (1)$$

the point here is to follow the well-known-wave procedure by making two partial derivatives of this expression with respect to space and one partial derivative with respect to time and by equating them we obtain the complex Schrödinger Wave Equation. But at this point it is important to recall the isomorphic property of Euler Relation as the one that makes it possible this result, as it were, the permanence of the unstable.

The first partial derivative with respect to space gives

$$\partial \Psi(x,t)/\partial x = J(2\pi/h)*p * \Psi * e^{2\pi J*p/h*x} * e^{-2\pi J/h*(p^2/2m)*t}$$

the internal derivative of

$$e^{2\pi J*p/h*x}$$

is $J(2\pi/h)*p$ and the rest of the expression can be represented as $\Psi''$ so

$$\Psi'' = \Psi * e^{2\pi J*p/h*x} * e^{-2\pi J/h*(p^2/2m)*t}$$

the second partial derivative gives us

$$\partial^2 \Psi(x,t)/\partial^2 x = (J(2\pi/h)*p)^2 * \Psi''$$

or else $\Psi''$

$$\Psi'' = \partial^2 \Psi(x,t)/\partial^2 x * 1/(J(2\pi/h)*p)^2$$

$$\Psi'' = -\partial^2 \Psi(x,t)/\partial^2 x * h^2/4\pi^2 * p^2 \qquad (2)$$

If we now take the partial derivative of (1) with respect to time, the internal derivative is $-2\pi J/h*(p^2/2m)$, so we have

$$\partial \Psi(x,t)/\partial t = -2\pi J/h*(p^2/2m)*\Psi''$$

or else

$$\Psi'' = \partial \Psi(x,t)/\partial t / -2\pi J/h*(p^2/2m)$$

$$\Psi'' = -\partial \Psi(x,t)/\partial t *(h*2mh/2\pi J*p^2) \qquad (3)$$

equating both $\Psi''$ expressions 2 and 3

$$\partial^2 \Psi(x,t)/\partial^2 x * h^2/4\pi^2 * p^2 = \partial \Psi(x,t)/\partial t *(h*2mh/2\pi J*p^2)$$

by eliminating and organizing terms of both sides we obtain finally:

$$\frac{\partial^2 \Psi(x,t)}{\partial^2 x} * \frac{h^2}{8\pi^2 * m} = \frac{h}{2\pi J} * \frac{\partial \Psi(x,t)}{\partial t}$$

we obtain the well-known Schrödinger Wave Equation, introduced by him in 1926, for a free particle moving in x's direction. This equation was presented then, as we said previously, as a postulate as there was no means to deduce it, from more basic principles, none the less, we have just applied the wave procedure to the BUS concept based on the principle of synergy.

There is an emergent conclusion in all this, and it is that *the fundamental of physical reality is energy*, and not a particle( or the mass concept). **A particle is a Bus or Holon rotating at an unknown low frequency but in the complex plane** or in reality represented in that plane, so that its state is completely determined, or else the qualitative aspect is reduced to the quantitative one in such a way no possibility of a field or just a storing capacity is feasible different from its well determined state. It is important to remember at this point that the Bus concept as a mathematical symbolism also has that wholeness attribute that is at the base of growth, each whole has the capacity to generate another whole, but always having in mind that openness attribute that also permits us to define the Bus in general as a symbol for representing a steady state open system. It is not a metaphysical concept that came from nothing. It is a concept in which a glueing principle as that of the Principle of Synergy permits us to define that oneness attribute.

## Conclusions and Suggested applications

From all this exposition emerges a conceptual framework that not necessarily reduces everything to the most elementary levels of reality, but as a good engineering conceptual tool opens new avenues for future research and development. The important point to recall at the outset is the need to abandon the old dualistic framework that has the natural tendency to put the whole or the part as a "primacy". The Bus concept is a part/whole complex mathematical concept that has embedded, as we have seen, the nondual and the dual nature of reality, but also the Principle of Synergy in which the whole is greater than the sum of its parts that gives us a medium to interpret reality, as it were, in organic ways as we have a minimum threshold of complexity.

In general, reality can be analyzed as a web of Buses but the Principle of Synergy implies a *same frequency* to obtain that "merry-go-around" effect or that emergent steady state or new order or that higher complex state where we have an organized complexity. One of the most important point to recall of this symbolism is that we do not need to make any reference to anthropomorphic concepts such as psyche or consciousness to explain those emergent states where we find a whole greater than the sum of the parts. But from the point of view of the whole reality we obtain a framework in which by the introduction of a fifth sphere of reality, that of form, we have then a sevenfold structure where the space-time continuum is embedded in that fifth sphere where life can be defined as an animated form, but then we have mind or the noosphere as the sixth sphere of reality, but also the Being as that one that has embedded them all.

Having found such a powerful explaining tool, it is obvious to feel an imperious need to share with the scientific community such a framework that can not only illuminates its actual practice by defining the objects of study in certain cases but also establishing clearly the impossibility to reduce those objects to isolable units in other cases. The whole concern of Ludwing von Bertalanffy in its GST when he wrote

*We may state as characteristic of modern science that this scheme of isolable units acting in one-way causality has proved to be insufficient. Hence the appearance, in all fields of science, of notions like wholeness, holistic, organismic, gestalt, etc., which all signify that, in the last resort, we must think in terms of systems of elements in mutual interaction.*

is then the same concern of all this paper, but our main point is the integration of those three big three, Being, Mind and Form or as Ken Wilber wrote in its **Sex, Ecology and Spirituality the Spirit of Evolution**

*With Kant, each of these spheres is differentiated and set free to develop its own potentials without violence...These three spheres, we have seen, refer in general to the dimensions of "it", of "we", and of "I"...In the realm of "itness" or empiric-scientific truths, we want to know if propositions more or less accurately match the facts as disclosed...In the realm of "I-ness", the criterion is **sincerity**...And in the realm of "we-ness" the criterion is goodness, or justness or relational care and concern...What is required, of course, is not a retreat to a predifferentiated state...what is required in the **integration of the Big Three**. And that, indeed, is what might be called the **central problem of postmodernity**...how does one integrate them?*

Science, philosophy and ontology and its integration implies a science that looks for truth but with an *openness* criterion, a philosophy than looks for *oneness* without the reductionism tendency but also for an ontology that looks for *wholeness* as a fundamental principle.

To manage complexity properly it is very essential to have a basic structure and among the possible conceptual structures we can have

- a binary or dual one and

- a threeness one that by mathematical inevitability becomes a sevenfold structure

This sevenfold structure has been used succesfully in many fields in a natural way but also recently we can find its application in many other fields too. For example the three-"tiers" architectural approach to client/server solutions and which looks for separating the various components of a client/server system into three "tiers" of services that must come together to create an application is precisely a solution whose main aim is to manage the changing complexity and which requires a basic hierarchy that starts with the service to the client. A tool must be adapted in every case so that we can avoid what can be named the "Galileo syncretic whole reasoning" about that sevenfold structure. Maybe Galileo powerful mind "saw" the powerful explaining capacity of this structure but his time was just the beginning of a science in which the bynary or dual structure would prevail.

Structure and environment are the main starting point of the new sciences of complexity or in the studies of complex adaptive systems in which adaptation to the environment implies always a minimum conceptual structural framework.

*User services* or interface or environment, ***business services*** or an adaptive plan, and ***data services*** or representation of changing estructures in a general way, are just some of the new applications we can found in the General Systems Sciences or Sciences of complexity of this sevenfold framework with which we definitely transcend the dualistic worldview and which is really very different from the holistic paradigm, and not just from the conceptual point of view but even more important from the practical point of view.

1. Bergson Henry.Creative Evolution. University Press of America.1983.

2. Bertalanffy von Ludwig. General System Theory George Braziller.1969.

3. Beer and Johnston. Vector Mechanics for Engineers. McGraw-Hill. 1962

4. Chardin Teilhard de. The Phenomenon of Man. Perennial Library.1975

5. Einstein et All. The Principle Of relativity. Dover.1952

6. Epsilon Pi. [Physics and The Principle of Synergy](#) Amazon.com.1999

7. Hazen and Pidd. FISICA. Editorial Norma. 1965

8. Ken Wilber. Sex, Ecology, Spirituality. Shambhala. 1995

9. Paternina Edgar. Física y Realidad. 1991.

10. Saussure Ferdinand de. Course in General Linguistic. 1997

11. Stuart Kauffman. At Home in the Universe. Oxford Paperbacks.1995.

12. Lao Tse. Tao Te King. Fontana. 1994

13. The I Ching. Richard Wilhelm Translation. Princenton .University Press.1990

14. Thomas S. Kuhn. The Structure of Scientific Revolutions. The University of Chicago Press.1996